\documentclass[fleqn,usenatbib]{mnras}

\usepackage{newtxtext,newtxmath}
\usepackage[T1]{fontenc}

\DeclareRobustCommand{\VAN}[3]{#2}
\let\VANthebibliography\thebibliography
\def\thebibliography{\DeclareRobustCommand{\VAN}[3]{##3}\VANthebibliography}

\usepackage{graphicx}	% Including figure files
\usepackage{amsmath}	% Advanced maths commands

\usepackage{siunitx}    % For formatting units
\DeclareSIUnit\Mpc{Mpc}
\DeclareSIUnit\Myr{Myr}
\DeclareSIUnit\Gpc{Gpc}
\DeclareSIUnit\h{\ensuremath{\mathit{h}}}
\DeclareSIUnit\hrs{hrs}

\usepackage{enumitem}
\usepackage{scalefnt}
\usepackage{soul}

\usepackage{bm}

\usepackage{tikz}
\usepackage{pgfplots}
\usepackage{subfig}

\newcommand{\HI}{\textsc{H{\kern0.08334em}I}}
\newcommand{\HII}{\textsc{H{\kern0.08334em}II}}
\newcommand{\xII}{x_\text{II}}
\newcommand{\Nexus}{\textsc{NEXUS}}
\newcommand{\nexus}{\textsc{NEXUS+}}
\newcommand{\logFilter}{Log-Gaussian}

\newcommand{\TOcmFast}{{\emph{{21cmFAST}}}}
\newcommand{\TOcmSense}{{\emph{{21cmSENSE}}}}

\title[Persistent topology of reionization II]{Persistent topology of the reionization bubble network. II: \ \ Evolution \& Classification}

\author[Elbers \& van de Weygaert]{Willem Elbers$^{1}$ and Rien van de Weygaert$^{2}$\\
$^{1}$Institute for Computational Cosmology, Department of Physics, Durham University, South Road, Durham, DH1 3LE, UK,\\
$^{2}$Kapteyn Astronomical Institute, University of Groningen, P.O. Box 800, 9700AV Groningen, the Netherlands}

\date{Last updated 18 June, 2022; in original form 18 June, 2022}

\pubyear{2022}

\pubyear{2022}

\definecolor{lightgrey}{RGB}{230, 230, 230}

\begin{document}
\label{firstpage}
\pagerange{\pageref{firstpage}--\pageref{lastpage}}
\maketitle

% Abstract of the paper
\begin{abstract}
We study the topology of the network of ionized and neutral regions that characterized the intergalactic medium during the Epoch of Reionization. Our analysis uses the formalism of persistent homology, which offers a highly intuitive and comprehensive description of the ionization topology in terms of the births and deaths of topological features. Features are identified as $k$-dimensional holes in the ionization bubble network, whose abundance is given by the $k$th Betti number: $\beta_0$ for ionized bubbles, $\beta_1$ for tunnels, and $\beta_2$ for neutral islands. Using semi-numerical models of reionization, we investigate the dependence on the properties of sources and sinks of ionizing radiation. Of all topological features, we find that the tunnels dominate during reionization and that their number is easiest to observe and most sensitive to the astrophysical parameters of interest, such as the gas fraction and halo mass necessary for star formation. Seen as a phase transition, the importance of the tunnels can be explained by the entanglement of two percolating clusters and the fact that higher-dimensional features arise when lower-dimensional features link together. We also study the relation between the morphological components of the bubble network (bubbles, tunnels, islands) and those of the cosmic web (clusters, filaments, voids), describing a correspondence between the $k$-dimensional features of both. Finally, we apply the formalism to mock observations of the 21-cm signal. Assuming 1000 observation hours with HERA Phase II, we show that astrophysical models can be differentiated and confirm that persistent homology provides additional information beyond the power spectrum.
\end{abstract}

% Select between one and six entries from the list of approved keywords.
% Don't make up new ones.
\begin{keywords}
cosmology: theory -- dark ages, reionization, first stars -- large-scale structure of Universe -- intergalactic medium\vspace{-1em}
\end{keywords}

%%%%%%%%%%%%%%%%%%%%%%%%%%%%%%%%%%%%%%%%%%%%%%%%%%

%%%%%%%%%%%%%%%%% BODY OF PAPER %%%%%%%%%%%%%%%%%%

\section{Introduction}

The anticipated detection of a 21-cm signal from the Epoch of Reionization will be an important milestone in the development of intensity mapping, an ambitious effort to map the three-dimensional distribution of gas in the Universe through the redshifted spectral line of neutral hydrogen. By measuring spatial variations in the intensity of the 21-cm signal, radio interferometers such as LOFAR \citep{vanhaarlem13}, MWA \citep{tingay13}, HERA \citep{deboer17}, and SKA \citep{dewdney09} could survey a volume far exceeding that which is currently accessible, with enormous potential for cosmology and astrophysics \citep{furlanetto06b,pritchard12,liu20}. During the Dark Ages ($z\gtrsim30$), the signal is a clean tracer of the total matter distribution, permitting novel constraints on the primordial power spectrum and fundamental physics. At lower redshifts, the signal depends on the temperature, density, and ionization fraction of hydrogen, revealing the state of the intergalactic medium (IGM) as it evolved through the Cosmic Dawn and Epoch of Reionization, the periods between $5.5\lesssim z\lesssim 20$ when the first stars were formed and feedback processes heated and subsequently ionized the IGM.

Current observations of the Cosmic Dawn and Epoch of Reionization are limited, covering only a small fraction of the total volume. For instance, polarization measurements of the Cosmic Microwave Background only probe the integrated optical depth \citep{planck20} and for the Lyman-$\alpha$ forest one is restricted to the available sight-lines \citep{becker15,bosman18,eilers18,garaldi19,yang20}. In terms of our understanding of these early epochs, the spatial and tomographic information encoded in the 21-cm signal would be transformational. However, untangling this signal remains challenging due to its weakness, the presence of bright foregrounds, and the size of the astrophysical model space. Currently, only upper limits on the 21-cm power spectrum exist \citep{mertens20,trott20,abdurashidova22} and the creation of three-dimensional maps that capture the rich topology of the signal will require an even greater sensitivity. This calls for a data analysis framework capable of extracting interpretable information from noisy channels. In this paper, which is a continuation of our previous work (\citealt{whe_2018a}; henceforth Paper I), we argue that persistent homology offers such a framework, and one that is ideally suited for reionization due to its grounding in topology.

It has long been recognized that topology provides a salient description of the spatial inhomogeneity of reionization \citep{gnedin00,lee08}. In the topological picture, we follow the spatial connectivity of the network of ionized regions to characterize the process of reionization. A basic description of this process proceeds as follows. During the early stages, isolated \HII{} bubbles form around the first sources. As the ionization front advances, these bubbles link up to form connected regions of ionized material, which encircle tunnels of neutral gas. Eventually, the openings of these tunnels are ionized as well, creating cavities that contain the last remaining islands of neutral material. These cavities are finally ionized from the outside in. Following Paper I, we call this growing structure the `ionization bubble network', although a more appropriate name would also reflect the role of the neutral regions and the prominent tunnels that connect them. Among these features, the tunnels are of particular importance, as they relate to the percolation processes associated with reionization \citep{furlanetto16a,bag18}. We will show that the peak of their prominence coincides with the Epoch of Reionization proper.

The standard summary statistic for 21-cm fluctuations is the power spectrum \citep{furlanetto04c,zaldarriaga04,mellema06,mcquinn07,pober14}, but non-Gaussianity implies that complementary observables contain additional information (we will confirm this explicitly for persistent homology in Section \ref{sec:thermal}). A range of such statistics have been proposed in the literature, including the bispectrum \citep{shimabukuro17,majumdar18,hutter21,watkinson22}, Minkowski functionals \citep{gleser06,friedrich11,yoshiura16,kapahtia18,bag18,chen19}, and the size distribution of bubbles \citep{lin16,giri17,kakiichi17,bag18}\footnote{See Paper I and references therein for other examples.}. In this paper, we study the 21-cm signal using the theory of persistent homology. This formalism offers a highly intuitive and comprehensive description of the ionization topology in terms of the births and deaths of topological features (components, tunnels, and cavities). A notable advantage of the framework is its ability to quantify the significance of topological features, which sets it apart from global quantities like the genus and Betti numbers (see Section \ref{sec:pershom}). This is particularly useful for extracting genuine astrophysical features from noisy observations, but also for uncovering the multiscale nature of the network that arises from the hierarchical build-up of structure. Another key advantage is its ability to identify tunnels in the bubble network. As mentioned above, tunnels are an important tracer of reionization, but one for which the power spectrum is ill-suited due to its lack of sensitivity to one-dimensional filamentary structures \citep{obreschkow12}.

In Paper I, we already gave an extensive description of the theory of persistent homology and used it to study a number of phenomenological models of reionization. We identified different stages of reionization based on the types of features that dominate. In this work, we apply the formalism to realistic mock observations generated with the semi-numerical code \TOcmFast{} \citep{murray20}. Our goal in this paper is twofold: (i) to study the evolution and persistence of the ionization topology in more realistic scenarios and (ii) to test whether persistent homology can be used to extract astrophysical information from mock observations.

The remainder of the paper is structured as follows. In Section \ref{sec:methods}, we briefly describe our methods and review some essential elements of the formalism. In Section \ref{sec:sims}, we describe our simulations and pipeline for including observational effects. We then focus on the evolution and structure of the ionization topology, describing the evolution with redshift in Section \ref{sec:evolution} and discussing the link with the topology of the cosmic web in Section \ref{sec:cosmic_web}. We then apply the formalism to the thermal structure of the 21-cm signal itself and describe its use as a classification tool in Section \ref{sec:thermal}. Finally, we provide a discussion and concluding remarks in Section \ref{sec:discussion}.

\subsection{Persistent Homology}\label{sec:pershom}

Topology is the study of properties that are conserved under continuous deformations, such as bending or stretching. One of the most elementary such properties is the genus $g$, simply put the number of holes in a surface. The genus has been widely applied in the context of reionization (e.g. \citealt{gleser06,lee08,friedrich11,hong14}). The notion of holes can be generalized, leading to the definition of Betti numbers. Informally, the $k$th Betti number gives the number of $k$-dimensional holes. For three-dimensional objects, there are three relevant numbers: $\beta_0$ describes the number gaps or connected components, $\beta_1$ the number of openings or tunnels, and $\beta_2$ the number of cavities or shells. Applying these concepts to the ionization bubble network, we find that $\beta_0$ describes the number of ionized regions, $\beta_1$ the number of neutral or ionized tunnels, and $\beta_2$ the number of enclosed neutral patches. We collectively refer to the bubbles, tunnels, and patches as topological features.

In algebraic topology, $\beta_k$ is the rank of the $k$th homology group, itself an algebraic representation of the $k$-dimensional holes \citep{hatcher01,edelsbrunner10b,carlsson21}. The Betti numbers are related to the more familiar Euler characteristic via the alternating sum
\begin{align}
\chi = \beta_0 - \beta_1 + \beta_2. \label{eq:chi}
\end{align}

\begin{figure}
	\subfloat{
		\hspace{-3em}
		\includegraphics{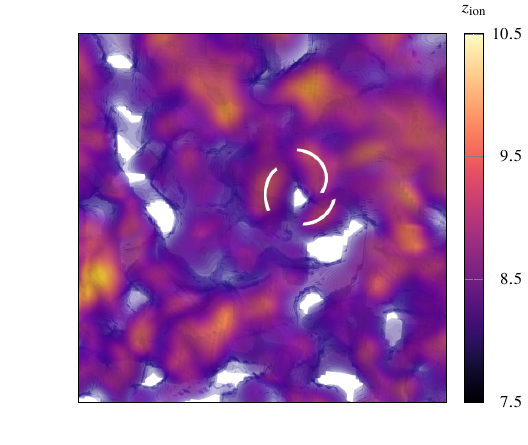}
	}\vspace{-2em}
	\caption{Ionized regions at $z=7.5$ in a cuboid \SI{80}{\Mpc} deep and \SI{300}{\Mpc} on a side, smoothed with a Gaussian filter with a $\rm{FWHM}$ of $\SI{15}{\Mpc}$. The regions are coloured by the redshift of first ionization $z_\text{ion}$, according to the `Faint Galaxies' model introduced in Section \ref{sec:21cmfast}. The bright spots correspond to early \HII{} bubbles. Three such regions have been marked to indicate that tunnels are formed once bubbles link up.}
	\label{fig:3d_image}
\end{figure}

\noindent
Evidently, the Betti numbers contain strictly more information than the Euler characteristic or genus. However, we can go further by keeping track of individual features as some underlying parameter $\alpha$ is varied. We assign every feature a pair of numbers ($\alpha_\text{birth},\alpha_\text{death})$, corresponding to the values at which the feature appears and disappears. The persistence of a feature is the difference $\alpha_\text{death}-\alpha_\text{birth}$ \citep{edelsbrunner00,zomorodian05}. For each dimension $k$, there exists a persistence diagram representing the set of $k$-dimensional features in $(\alpha_\text{birth},\alpha_\text{death})$-space. The Betti numbers $\beta_k(\alpha)$ can be reconstructed from the persistence diagrams as a function of $\alpha$. For example, we can follow the evolution of the ionization topology as a function of time $t$. The number of components $\beta_0$ increases by 1 when a bubble is born and decreases by 1 when two regions merge, and similarly for the tunnels and neutral patches. The diagrams for $\beta_k$ and $\beta_{k+1}$ are furthermore related, since higher-dimensional features arise when lower-dimensional features link together (see Fig.~\ref{fig:3d_image}). Persistence is useful as a measure of topological significance: features that exist only within a narrow interval $[\alpha,\alpha+\epsilon]$ are less significant and more likely to be noise than features that are extremely persistent.

Besides our earlier work (\citealt{elbers17}, Paper I), Betti numbers have been used in the context of reionization by \citet{kapahtia18,kapahtia19,kapahtia21,giri20} and \citet{bianco21}. Among these, the work of \citet{giri20} is most closely related to our own, while \citet{kapahtia18,kapahtia19,kapahtia21} analyse two-dimensional temperature maps. An important difference with these works is that our analysis accounts for the persistence of features. This allows us to quantify their significance, which is a useful analytical tool and crucial for applications to low signal-to-noise maps. Persistence was also used by \citet{thelie22} to identify significant ionized patches, following a different but related formalism based on Morse theory. In this work, we will analyse the full topology of the bubble network, including its tunnels and neutral patches, by deriving persistence diagrams for all three dimensions.

In recent years, persistent homology has become a popular tool in cosmology due to its ability to capture the complex multiscale topology that arises from nonlinear structure formation and identify its most significant features. The most fruitful applications have been in studies of the cosmic web \citep{weygaert11,sousbie11,pranav16,xu18,wilding21,bermejo22}, which have shown that the persistent homology of the cosmic density field reflects the hierarchical build-up of structure, and to the Gaussianity of random fields \citep{feldbrugge12,park13,cole18,feldbrugge19,cole20,biagetti20}. The versatility of the formalism is reflected by other wide-ranging applications, including most recently to interstellar magnetic fields \citep{makarenko18} and baryon acoustic oscillations \citep{kono20}. The formalism can be used to improve constraints on cosmological parameters, as demonstrated effectively by \citet{heydenreich20} in the case of cosmic shear.

\section{Methods}\label{sec:methods}

Here, we briefly describe our methods for calculating topological statistics and review some essential elements of the formalism. For a detailed discussion of the theory, we refer to Paper I.

\subsection{Field Filtrations}

We use the Field Filtration method to describe the ionization topology as a function of different parameters. The method works by taking superlevel sets of three-dimensional scalar fields. In this paper, we consider two different filtration parameters\footnote{Other filtration parameters were proposed in Paper I, such as the ionization fraction $\xII$ and spatial course-graining scale $\alpha$.}, each bringing to light a different aspect of the ionization topology.

\begin{enumerate}[label=(\roman*), wide=0pt, widest=99,leftmargin=\parindent, labelsep=*]
	\item Redshift $z$. In the first part, we consider the evolution of the ionization topology with redshift, offering a formal description of the ionization process. This description reveals a multiscale organisation that reflects the hierarchical evolution of the underlying cosmic density field. The three-dimensional field is the $z_\text{ion}$ field, giving the redshift of first ionization of each cell. The persistence of a feature represents its lifetime.
	\item Temperature $T_b$. In the second part, we apply the formalism directly to mock observations and show that we can distinguish astrophysical models based on the topology. In this case, the three-dimensional field is the 21-cm temperature field and the persistence of a feature represents the range of temperatures for which it exists. This allows genuine features to be distinguished from thermal fluctuations.
\end{enumerate}

After the choice of parameter has been made, we construct a filtration that captures the topology of the associated three-dimensional field $f(\mathbf{x})$. The filtration is a nested sequence of objects, called \emph{simplicial complexes}, constructed by taking superlevel sets of $f(\mathbf{x})$. A simplicial complex is a structure that is convenient for computational purposes and built from simplices: points, lines, triangles, and tetrahedra. The filtration is constructed as follows. We start by computing a periodic Delaunay triangulation of the grid on which the field values are given. This represents the final `completed' simplicial complex. A vertex $v$ from the triangulation is added to the filtration when the filtration parameter $\alpha$ exceeds the field value $f(v)$. Any higher-dimensional simplex is added at the lowest value of $\alpha$ for which each of its vertices are present. The complex is built with the computer package \textsc{cgal} \citep{cgal16} and its topology is computed with the \textsc{gudhi} library \citep{gudhi14}.

The Field Filtration method may be compared to the integral geometric approach used to compute Minkowski functionals \citep{mecke94,schmalzing97}. In both cases, the structure of the field is studied using superlevel sets. The Minkowski functionals describe the geometry of the superlevel sets in terms of the volume, surface area, and mean curvature, as well as the topology, through the Gauss-Bonnet theorem, in terms of the Euler characteristic. Similarly, the Betti numbers and persistence diagrams computed with the Field Filtration method describe the topology of the superlevel sets in terms of $d$-dimensional feature counts. As noted before, the Betti numbers are related to the Euler characteristic through the alternating sum \eqref{eq:chi}, but contain additional information. As such, the integral geometric and Field Filtration methods are similar but complementary.

\begin{figure*}
	\subfloat{
		\includegraphics{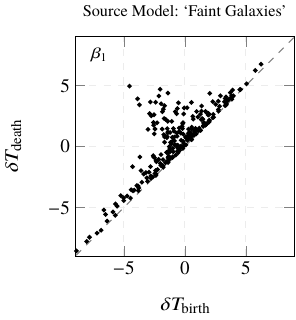}
	}
	\subfloat{
		\includegraphics{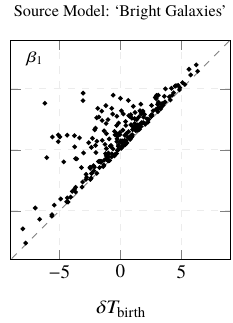}
	}
	\subfloat{
		\includegraphics{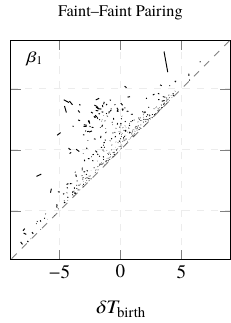}
	}
	\subfloat{
		\includegraphics{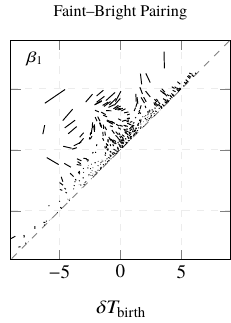}
	}
	\caption{(Left) Persistence diagrams, showing the births and deaths -- and overall significance -- of tunnels in the ionization bubble network. Shown are persistence diagrams for two reionization source models: `Faint Galaxies' and `Bright Galaxies', introduced in Section \ref{sec:21cmfast} overleaf. Features with greater persistence, $\delta T_\text{death} - \delta T_\text{birth}$, are more significant and lie further from the diagonal. (Right) Optimal pairings between persistence diagrams. The first plot represents a pairing between different random realizations of the same astrophysical model. The second plot represents a pairing between realizations of different models. Evidently, the total distance is much larger between simulations with different models.}
	\label{fig:distance_example}
\end{figure*}

\subsection{Persistence diagrams}

A persistence diagram is a plot of features in $(\alpha_\text{birth},\alpha_\text{death})$-space. The advantage of persistence diagrams is that they allow us to differentiate between significant topological features that exist over a wide range of scales and insignificant features that are more likely to be noise. Significant features will have a larger persistence, $
\alpha_\text{death}-\alpha_\text{birth}$, and lie further away from the diagonal, as shown in Fig.~\ref{fig:distance_example}. Given two persistence diagrams $X$ and $Y$, we can form a one-to-one correspondence $\phi\colon X\to Y$ that matches each point in $X$ with a point in $Y$ and vice versa. Each diagram contains infinitely many copies of the diagonal, which we treat as a point $\Delta$ that can be matched with another copy of the diagonal or with an off-diagonal point. In Fig.~\ref{fig:distance_example}, we show two examples of such pairings for diagrams of one-dimensional features (tunnels). The pairing on the left is between two random realizations of the same reionization source model (`Faint Galaxies'). On the right, we show a pairing with a realization of a different model (`Bright Galaxies'). The details of these models are given in Section \ref{sec:21cmfast}.

% \rwr{A bit more explanation that this is about significance of diagrams.}

Given a pairing $\phi$, we can compute the Euclidean distance between any two matched points $\rvert\!\rvert x -\phi(x)\rvert\!\rvert$. A pairing is said to be \emph{optimal} if it minimises the total squared distance between all points. The total $L^2$-Wasserstein distance between the diagrams is then defined as \citep{turner14,boisonnat18}
\begin{align}
	d(X,Y) = \left[\inf_{\phi\colon X\to Y}\sum_{x\in X}\rvert\!\rvert x -\phi(x)\rvert\!\rvert^2\right]^{1/2}. \label{eq:wasserstein}
\end{align}

\noindent
In other words, the distance between two diagrams is the square root of the total squared distance of an optimal pairing. The pairings in Fig.~\ref{fig:distance_example} are optimal. Evidently, the distance between diagrams from the same astrophysical model is much less than the distance between diagrams of different models. Moreover, the distance is dominated by high-persistence features. Noisy features close to the diagonal can always be matched with $\Delta$ and therefore have a negligible impact on the Wasserstein distance. We will exploit this property to differentiate between noisy observations from different models.

To enable a statistical description, we also define summary statistics for samples of diagrams. A set $\{X_i\}$ of $n$ persistence diagrams define a Fr\'echet function
\begin{align}
    F(Y) = \frac{1}{n}\sum_{i=1}^nd(Y,X_i)^2.
\end{align}

\noindent
A \emph{Fr\'echet average} of $\{X_i\}$ is a diagram $Y$ that minimizes $F(Y)$. The \emph{Fr\'echet variance} of $\{X_i\}$ is the minimum $F(Y)$. See the discussion surrounding Eqns. (5--7) in paper I for more details. These summary statistics can be combined into a visual representation called a \emph{persistence field}. Persistence fields resemble persistence diagrams, but also reflect the density and statistical uncertainty of features. All smoothed persistence diagrams shown in this paper are persistence fields and we will use the terms synonymously.

\section{Simulations}\label{sec:sims}

We apply our formalism to realistic simulations of the ionization field and the resulting 21-cm signal. In the first part of the paper, we study the evolution of the ionization topology assuming perfect knowledge of the ionization state of the IGM. In the second part, we apply the formalism to mock observations of the 21-cm differential brightness temperature. The simulations are run with the semi-numerical code \TOcmFast{}, described in Section \ref{sec:21cmfast}. Our treatment of instrumental effects is described in Section \ref{sec:instrumentals}.

\subsection{Reionization simulations}\label{sec:21cmfast}

We make use of \TOcmFast{} \citep{murray20}, a semi-numerical code based on excursion set principles \citep{mesinger07,mesinger11}. The basic operation is as follows. First, an initial Gaussian density perturbation is generated on a grid, which is then evolved forward in time using Lagrangian perturbation theory. Ionized regions are identified using the excursion set formalism. On a courser grid, the number of ionizing photons within spheres of decreasing radius is compared with the number of recombinations to determine whether a cell is ionized.

In determining the number of ionizing photons, the galactic UV radiation is calculated from the mean baryon density in each sphere using the prescription of \citet{park19}. In this prescription, the number of ionizing photons per baryon $\zeta=f_*f_\text{esc}N_{\gamma/b}$ is broken up into parts: the fraction $f_*$ of gas contained in stars, the escape fraction of ionizing photons $f_\text{esc}$, and a normalisation factor $N_{\gamma/b}=5000$. The first two factors are assumed to follow a power law with respect to halo mass: $f_*=f_{*,10}M_h^{\alpha_*}$ and $f_\text{esc}=f_{\text{esc},10}M_h^{\alpha_\text{esc}}$, normalized at $10^{10}\;M_\odot$.

It is expected that X-ray sources heat the IGM prior to reionization taking off \citep{oh01,venkatesan01,ricotti04}. This process is modelled by calculating the intensity of X-ray radiation at each cell, from which the initial ionization fraction $\xII$ and spin temperature $T_S$ at each cell are computed. Finally, the 21-cm differential brightness temperature field $\delta T_b$ can be calculated using (e.g. \citealt{pritchard12})
\begin{align}
\delta T_b(z) = T_0(z)(1+\delta_b)(1-x_\text{II})\left(1-\frac{T_\text{CMB}(z)}{T_\text{S}}\right), \label{eq:delta_T}
\end{align}

\noindent
where $T_0(z)$ is a function of cosmological parameters and redshift, $\delta_b$ the baryonic overdensity, $T_\text{CMB}$ the CMB temperature, and $T_\text{S}$ the spin temperature. For more details, we refer to \citet{mesinger11,park19}.

We assume a standard flat $\Lambda\text{CDM}$ cosmology with $h=0.68$, $n_s=0.97$, $\sigma_8=0.81$, $\Omega_{\rm{m}}=0.31$, $\Omega_{\rm{b}}=0.0048$. Based on \citet{greig17}, we distinguish two astrophysical scenarios with contrasting reionization morphologies:

\begin{enumerate}[label=(\roman*), wide=0pt, widest=99,leftmargin=\parindent, labelsep=*]
	\item \emph{Faint galaxies:} $M_\text{turn}=0.5\times10^9\; M_\odot$,  $f_{*,10}=0.05$.
	\item \emph{Bright galaxies:} $M_\text{turn}=7.5\times10^9\; M_\odot$, $f_{*,10}=0.15$.
\end{enumerate}

The scenarios differ in two astrophysical parameters: $M_\text{turn}$, which is the minimum halo mass below which star formation is suppressed exponentially due to feedback, and $f_{*,10}$, which is the fraction of gas contained in stars normalised for halos with mass $10^{10}\,M_\odot$. We use the fiducial values of \citet{park19} for the remaining astrophysical parameters in the model: $\alpha_*=0.5$, $f_{\text{esc},10}=0.1$, $\alpha_\text{esc}=-0.5$, the star formation time-scale $t_*=0.5$ in units of Hubble time $H^{-1}$, and the minimum energy $E_0=\SI{0.5}{\keV}$ necessary for an X-ray to escape. The two scenarios were chosen to represent the likely range of reionization topologies, but with both scenarios achieving complete reionization at $z\sim6$. In addition to these two models, we also explore the impact of radiation sinks by running a model with fewer recombinations.

In the first part of the paper, we study the evolution of the ionization topology for these different scenarios. The persistence fields shown in the next section represent the Fr\'echet average of three realizations of a $(\SI{300}{\Mpc})^3$ cube, corresponding to the fiducial `Faint Galaxies' model. The density fields were evolved on $1024^3$ grids and the ionization fields and topology were calculated on $256^3$ grids. We evolved a single realization for each of the alternative models considered in this paper using the same grid and box sizes. In the second part, we use many smaller realizations of $(\SI{300}{\Mpc})^3$ cubes with just $512^3$ density grids and $128^3$ $\delta T_b$ grids.

\subsection{Instrumental effects}\label{sec:instrumentals}

The Hydrogen Epoch of Reionization Array (HERA) recently reported the first results from Phase I of the experiment \citep{abdurashidova22}, setting improved upper limits on the 21-cm power spectrum. In this paper, we will model instrumental effects based on 1000 hours of observation with Phase II of HERA \citep{deboer17}, following a procedure similar to that of \citet{hassan19}, but accounting for lightcone effects \citep{greig18}. We assume a 350-element layout, consisting of 320 elements tightly packed in a hexagonal core and 30 outlying elements. We deal with three main instrumental effects in order:

\begin{enumerate}[label=(\roman*), wide=0pt, widest=99,leftmargin=\parindent, labelsep=*]
    \item angular resolution of the instrument,
    \item foreground removal or avoidance,
    \item thermal noise.
\end{enumerate}

\subsubsection{Angular resolution}

Radio interferometers make observations in $uv$-space, which need to be transformed to comoving distances. Baseline lengths $\mathbf{u}=(u,v)$ are related to comoving wavenumbers $\mathbf{k}_\perp=(k_x,k_y)$ in the plane orthogonal to the line of sight according to \citep{furlanetto06b}
\begin{align}
\mathbf{k}_\perp = \frac{2\pi\mathbf{u}}{D_c(z)},
\end{align}

\noindent
where $D_c(z)$ is the comoving distance at redshift $z$. The longest baselines will determine the angular resolution of the instrument. We account for the redshift-dependence of the resolution by computing the intensity of baseline coverage at each $uv$ pixel at 10 redshifts between $6\leq z\leq25$ with \TOcmSense{} \citep{pober16}. Pixels with minimal $uv$-coverage, corresponding to outrigger baselines, contribute most of the thermal noise. We find that a cut-off of 20\% on the pixels with the least $uv$-coverage benefits topological inference, by reducing thermal noise at the cost of limiting the angular resolution. To apply the resolution to our lightcones, we compute the Fourier transform of each cubic slice along the redshift direction and discard modes with zero $uv$-coverage at the bounding redshifts. Finally, we linearly interpolate between the inverse Fourier transforms of the cubes along the redshift direction.

\subsubsection{Foreground removal}

Contamination by foreground emission is a major impediment to 21-cm observations of the EoR. This effect is mainly restricted to a wedge in Fourier space \citep{liu14,pober14},
\begin{align}
	k_\parallel \leq \frac{H(z) D_c(z)}{c(1+z)}\sin(\theta)\,k_\perp,
\end{align}

\noindent
where $k_\parallel$ is the wavenumber parallel to the line of sight, $H(z)$ the Hubble rate, and $\theta$ the angular radius of the field of view. The case with $\sin(\theta)=1$, known as the horizon limit, applies if foregrounds cannot be removed. This leaves a window in which the EoR can be observed relatively unobstructedly. We consider two possible scenarios following \citet{pober14}. In the moderate scenario, foreground emission bleeds into the EoR window affecting modes up to $k_\parallel=\SI{0.1}{\h\per\Mpc}$ beyond the horizon limit. We discard all modes below the horizon plus a $\SI{0.1}{\h\per\Mpc}$ buffer. The optimistic scenario of \citet{pober14} assumes that a successful foreground removal strategy can be found, such that only modes below the FWHM of the primary beam need to be discarded ($\theta<\text{FWHM}/2$). We compute both foreground models with \TOcmSense{} and apply them to the lightcones in the same way as the angular resolution.

\subsubsection{Thermal noise}

Thermal noise can be modelled as a Gaussian random field with noise power spectrum \citep{zaldarriaga04,pober14}
\begin{align}
	\Delta_N^2(k)  = X^2Y \frac{k^3}{2\pi^2}\frac{\Omega'}{2t}T_\text{sys}^2. \label{eq:noise_pow}
\end{align}
\noindent
where $X^2Y$ is a cosmological conversion factor, $t$ is the integration time for mode $k$, $\Omega'$ is a beam-dependent factor, and $T_\text{sys}$ is the system temperature. The system temperature is $T_\text{sys}=T_\text{sky} + T_\text{recv}$, where we adopt $T_\text{recv}=\SI{100}{K}$ and $T_\text{sky}=\SI{60}{\K}\,(\lambda/\SI{1}{\m})^{2.55}$ \citep{richard17}. We generate a Gaussian random field with power spectrum \eqref{eq:noise_pow} and divide the noise by the amount of $uv$-coverage at each $(k_\perp,k_\parallel)$ pixel. Given that $\Omega'\sim\lambda^2$, the overall redshift dependence is approximately $\Delta_N^2\sim (z+1)^{7.5}$. To properly include this redshift-dependence in our lightcones, we simulate noise cubes at 10 redshifts between $6\leq z\leq25$ and interpolate along the line of sight. We smooth the final signal in each cube with an isotropic Gaussian filter with smoothing radius $\SI{0.24}{\MHz}$, which corresponds to $\SI{4}{\Mpc}$ at $z\sim8$. We subtract the average temperature in each two-dimensional slice along the redshift direction, since absolute calibration is not possible. Before calculating the topology, we reduce the resolution further by shrinking the cubic slices to $64^3$ voxels in order to speed up the calculation.

Our main analysis assumes 1000 hours of observation. However, to investigate the effects of thermal noise, we also linearly scale the noise fields by factors between $0.1$ and $10$, corresponding to $10^5$ and $10^1$ hours of observation, respectively.

\begin{figure*}
	\centering
	\subfloat{
		\includegraphics{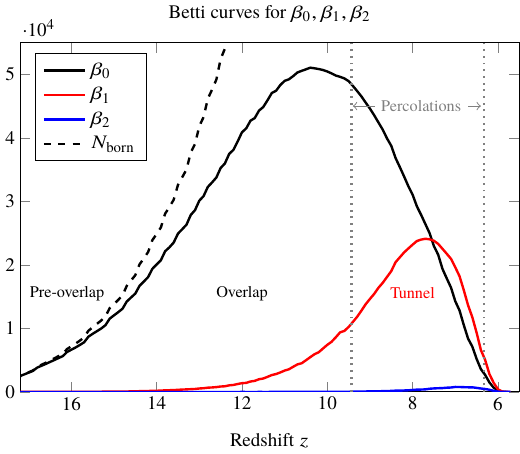}
	}
	\subfloat{
		\includegraphics{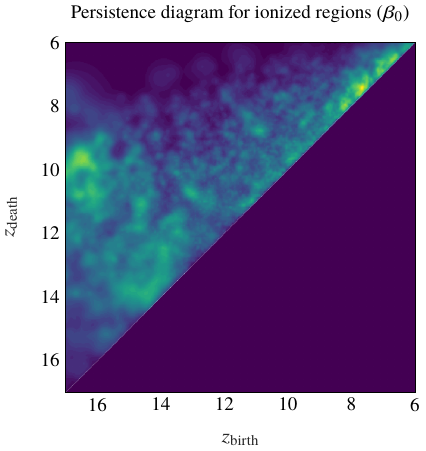}
	}
	\caption{(Left) Topological evolution of the fiducial `Faint Galaxies' model. Betti curves show the number of ionized regions, tunnels, and neutral regions alive as a function of redshift. The dashed line is the total number, $N_\text{born}$, of ionized regions that have been born. The vertical dotted lines indicate the two percolation transitions. (Right) The persistence diagram for $\beta_0$ shows the births and deaths of ionized regions.}
	\label{fig:betti_tracks}
\end{figure*}

\begin{figure*}\vspace{-2em}
  \centering
 \subfloat{
    \centering
	\includegraphics{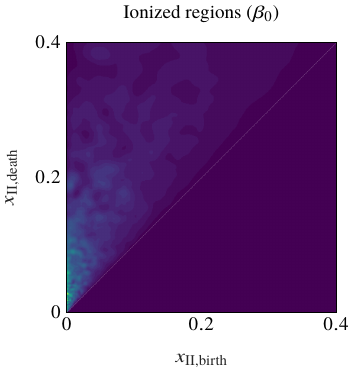}
  }\hspace{-2em}
   \subfloat{
    \centering
	\includegraphics{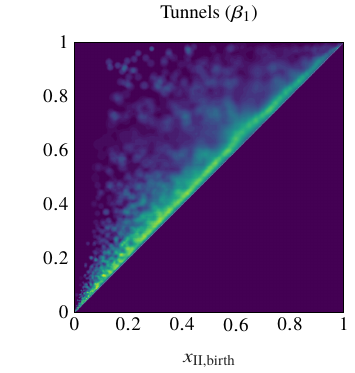}
  }\hspace{-2em}
  \subfloat{
    \centering
	\includegraphics{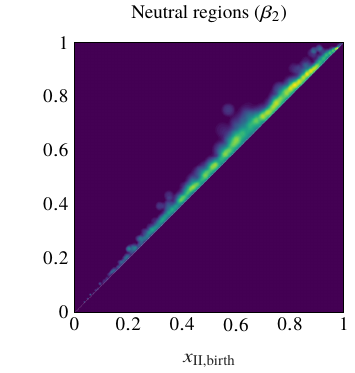}
  }
  \caption{Persistence diagrams for the `Faint Galaxies' model, mapped to ionization fraction, $\xII$, coordinates. Each point represents the birth and death of a topological feature in the ionization bubble network. The three diagrams are for features in dimensions $k=0,1,2$. Note that the axis limits are different for the first diagram $(\beta_0)$, as most of the pre-reionization evolution shown in the right panel of Fig.~\ref{fig:betti_tracks} is compressed in $\xII$-space.}
  \label{fig:iof_results}
\end{figure*}

\section{Topological evolution}\label{sec:evolution}

The first application of the formalism will be to give a theoretical description of the evolution of the ionization topology, without considering instrumental effects. This is done with a filtration of the $z_\text{ion}$ field, which gives the redshift of first ionization of each cell. The resulting Fig.~\ref{fig:betti_tracks} illustrates the main ideas of this section. The left panel shows Betti curves, describing the numbers of features alive at each redshift: ionized regions $\beta_0$ (black), ionized or neutral tunnels $\beta_1$ (red), and neutral regions $\beta_2$ (blue). The right panel is an example of a persistence diagram, in this case for $\beta_0$, showing the births and deaths of ionized regions in birth-death space. Further persistence diagrams, mapped to ionization fraction, $\xII$, coordinates are shown in Fig.~\ref{fig:iof_results}, now for all three dimensions.

In the remainder of this section, we describe how the topology evolves during each stage of heating and reionization (Section \ref{sec:global}), how the topology depends on the sources and sinks (Sections \ref{sec:sources} and \ref{sec:sinks}), as well as the spatial structure of the neutral regions (Section \ref{sec:spatial}). Finally, we discuss how percolation theory can explain the dominant role of the tunnels during the EoR (Section \ref{sec:perc_fil}),

\subsection{Global aspects}\label{sec:global}

First, let us broadly consider how the topology of the network evolves. This has traditionally been described in terms of the pre-overlap, overlap, and post-overlap stages of the ionization bubbles \citep{gnedin00}. These stages can be identified by taking the genus or Euler characteristic as a global indicator of the topology \citep{lee08,friedrich11,hong14,giri19}, but considering the Betti numbers separately, as we do here, allows one to identify additional stages during which the tunnels or neutral patches are important. Fig.~\ref{fig:betti_tracks} makes this abundantly clear. In the left panel, we show the Betti curves for the fiducial `Faint Galaxies' model. The $\beta_0$-curve (solid black) shows the number of ionized regions as a function of redshift $z$. Between $15<z<17$, this number increases gradually and mostly tracks the total number, $N_\text{born}$, of ionized regions that have been born (dashed black). Around $z=15$, the degree of overlap, $1-\beta_0/N_\text{born}$, reaches 10\%. This point marks the end of the pre-overlap stage, during which the topology was characterized by the emergence of distinct ionization bubbles.

Between $10<z<15$, $N_\text{born}$ increases rapidly as a younger generation of sources turn on. However, the number $\beta_0$ of distinct ionized regions reaches an inflection point at $z=12.5$ due to increased overlap. Ionized regions born at later times are less persistent, as shown in the persistence diagram in the right panel of Fig.~\ref{fig:betti_tracks}. We note that this remains true whether expressed in terms of redshift $z$ or proper time $t$. There are two physical reasons for the decreased persistence: younger bubbles arise in clusters and merge amongst themselves and they are more easily absorbed into pre-existing ionized structures. The pre-overlap and overlap stages cover the period $10<z<17$, corresponding to the epoch of IGM heating, which precedes the Epoch of Reionization. During this period, the ionization fraction, $\xII$, remains below a few per cent. As a result, one large neutral region dominates and higher-dimensional structures (such as tunnels or neutral islands surrounded by ionized material) are largely absent. Up to this point, the topology is well described by the single parameter $\beta_0$ and the size distribution of ionized regions gives an appropriate description of the geometry. Around $z=10$, just after $\beta_0$ reaches a maximum, a percolation transition occurs. This is a benchmark for the end of the overlap stage and the beginning of the `tunnel stage' (Paper I).

The number of bubbles decreases from $z=10$ onwards. The death of bubbles is associated with the birth of tunnels, which arise when ionized regions link up (as illustrated in Fig.~\ref{fig:3d_image}). Let us therefore consider the $\beta_1$-curve (red) for the number of tunnels in the network. After the first percolation transition, the tunnels become a significant component and they remain dominant throughout most of the reionization period. This is even more apparent when we look at the topology as a function of the ionization fraction, $\xII$, in the middle panel of Fig.~\ref{fig:iof_results}, which shows the continuous births of persistent tunnels until the end of reionization. The tunnels disappear rapidly following a second percolation transition around $z=6.5$. The persistence diagram for tunnels resembles a triangle in birth-death space. The two edges of the triangle correspond to the percolation transitions that bound the reionization era: the first transition is responsible for the vertical edge at $x_{\text{II},\text{birth}}\sim0.1$ and the second transition is responsible for the horizontal edge at $x_{\text{II},\text{death}}\sim0.95$. The two edges meet at an apex, marking the most significant tunnels present in the simulation, similar to that seen in persistence diagrams of the cosmic density field \citep{wilding21}. The dominance of the tunnels is best understood when we consider reionization as a percolation process (see Section \ref{sec:perc_fil}).

Finally, the number of neutral components is given by the $\beta_2$-curve (blue). These seem to be extremely rare, which is consistent with the findings of \citet{giri19,giri20} that neutral islands are much less common in the final stages of reionization than ionized regions are in the early stages. Indeed, Fig.~\ref{fig:betti_tracks} shows that the neutral regions never outnumber the tunnels. Since this was our criterion for the `neutral patch stage' (Paper I), it appears that a neutral patch stage is absent. This is slightly misleading, as we will we see in Section \ref{sec:spatial}, because significance in terms of number differs from  significance in terms of volume fraction. Nevertheless, the neutral regions are most numerous around the second percolation transition when the large neutral cluster breaks off into smaller neutral regions. The lack of persistence of the neutral regions (rightmost panel of Fig.~\ref{fig:iof_results}) is due to the fact that the neutral regions are quickly ionized once they break off from the percolating cluster.

\begin{figure*}
	\centering
	\includegraphics{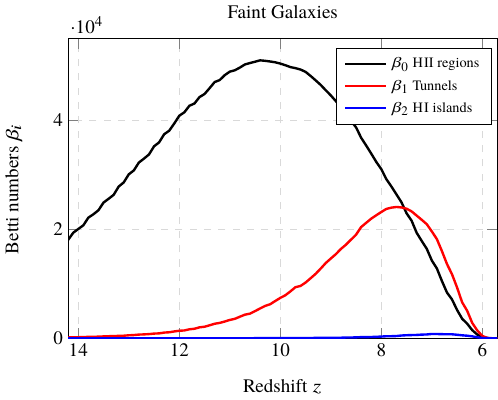}\hfill
	\includegraphics{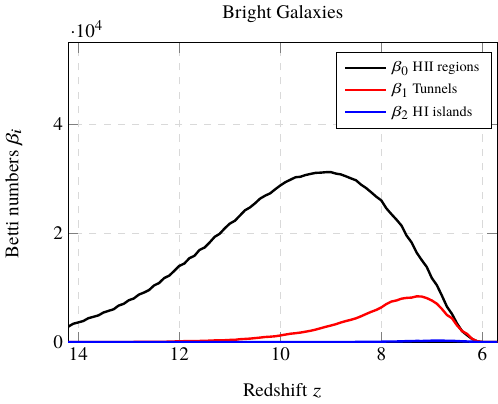}\\
	\includegraphics{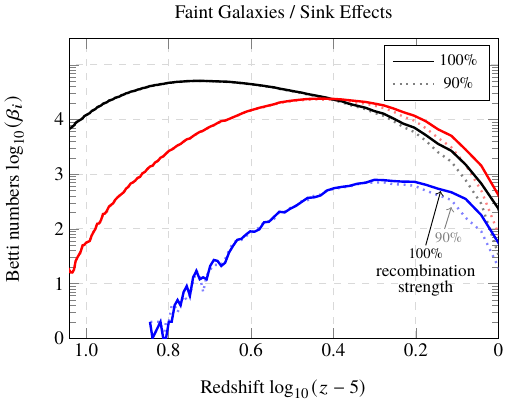}\hfill
	\includegraphics{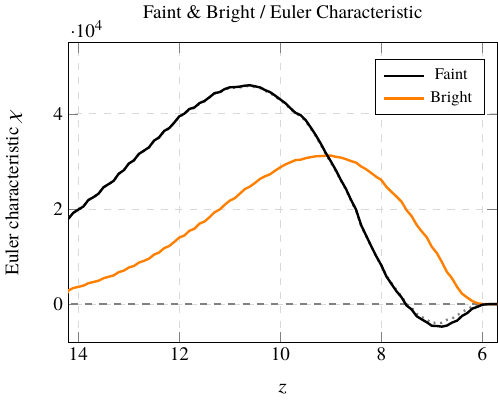}\\
	\caption{The effect of sources and sinks on the reionization topology shown in terms of Betti numbers. The top two panels compare two source populations: faint and bright galaxies. The bottom left panel shows the Faint Galaxies topology in log-log space, with the fainter dotted lines corresponding to a scenario with 10\% fewer recombinations. Note the logarithmic scaling of $z$ to emphasize differences around $z=6$. The bottom-right panel shows the Euler characteristic, $\chi=\beta_0-\beta_1+\beta_2$, for all three scenarios with the fainter dotted line again corresponding to the faint 90\% recombinations scenario.}
	\label{fig:sources}
\end{figure*}

\subsection{The effect of sources}\label{sec:sources}

In Fig.~\ref{fig:sources}, we compare the two source models in terms of their Betti curves. Both models are tuned to match observations and produce the same global ionization history that ends at $z=6$ \citep{greig17}, but the resulting $z_\text{ion}$ fields have markedly different topologies. The sources in the `Bright Galaxies' model are rarer and brighter. As a result, the ionized regions are larger but fewer in number, and we see fewer topological features of any kind. A second important effect is that the Epoch of Heating is delayed, due to the late formation of the sources.

The number $\beta_0$ of ionized regions in the `Bright Galaxies' model tracks the `Faint Galaxies' model within a few per cent after $z=8$. This means that, at the same ionization fraction, the connectivity and overlap of the large ionized regions that exist between $6<z<8$ are largely independent of the sources. This is because the global evolution during the EoR is similar for both models by construction. The main topological difference is in the neutral regions and especially the tunnels. This is encouraging, because we expect the tunnels to be the easiest to detect. It should be easier to estimate the number of holes than to identify whether there are any gaps between regions (which after all may connect out of view). Furthermore, $\beta_1$ can be measured by counting holes in either the neutral regions or the ionized regions. This prediction is confirmed when we apply the formalism to the 21-cm signal in Section \ref{sec:thermal}, at least when foregrounds can be successfully removed.

It is interesting to consider the Euler characteristic $\chi$ (bottom-right panel) as well. The Euler characteristic tracks the overall topological evolution described in Section \ref{sec:global} and can be used to distinguish different scenarios \citep{lee08,friedrich11,giri19}. However, considering the topological components independently reveals exactly why the Euler characteristic behaves as it does. Recalling Eqn. \eqref{eq:chi}, which relates $\chi$ to the Betti numbers, and the fact that neutral patches are rare, we learn that the evolution of the Euler characteristic mostly depends on the interplay between the number of ionized components and the number of tunnels. Before reionization, $\chi\approx\beta_0$ tracks the number of ionized regions, but during reionization $\chi\approx\beta_0-\beta_1$. The trough in the $\chi$-curve seen for the `Faint Galaxies' model corresponds to the Epoch of Reionization when the tunnels dominate. The depth of this trough is determined by the number of tunnels, but also by the relative timing of the Epoch of Heating and the Epoch of Reionization. By contrast, the EoH and EoR overlap in the `Bright Galaxies' model. The trough in the $\chi$-curve is absent for two reasons: the smaller numbers of topological features overall and the overlap between the $\beta_0$- and $\beta_1$-curves due to the delayed formation of the sources.

\subsection{The effect of sinks}\label{sec:sinks}

Next, we consider the effect of recombinations. We compare the fiducial `Faint Galaxies' model at 100\% recombinations with a `Faint Galaxies' model where the recombination coefficient is reduced by 10\%. This model represents a cosmology with a decreased abundance of absorbers such as Lyman limit systems. The results are shown in the bottom left panel of Fig.~\ref{fig:sources}. The number of features is largely unaffected at high redshift, but starts to deviate from $z=7.5$ onwards. For each dimension, the number of topological features is reduced by about 5\% in the 90\% recombinations model. This is due to the fact that ionizing photons can penetrate further, allowing the ionized network to expand uniformly in each direction compared to the fiducial model. While the decrease is similar in each dimension, the tunnels dominate during this time period, such that the effect of sinks is easiest to observe in the $\beta_1$-curve.

\begin{figure*}
	\subfloat{
		\centering
		\includegraphics{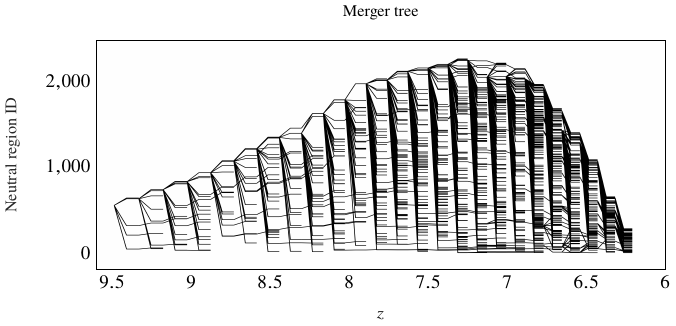}
	}
	\subfloat{
		\centering
		\includegraphics{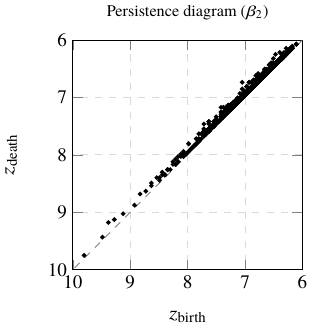}
	}
	\caption{(Left) The merger tree shows that most neutral regions disappear quickly after they split off from the percolating cluster, i.e. they have low persistence. (Right) This is easily seen in the persistence diagram for neutral islands $(\beta_2)$. Both diagrams are for the `Faint Galaxies' model.}
	\label{fig:merger_tree}
\end{figure*}

\subsection{Spatial structure of the neutral regions}\label{sec:spatial}

As discussed in Section \ref{sec:global}, enclosed neutral regions are born mostly during the second half of the EoR and are not very persistent. The information contained in a persistence diagram of dimension $d-1$ can also be represented as a merger tree. This is demonstrated by the two panels of Fig.~\ref{fig:merger_tree}. All neutral regions split off from the percolating cluster starting from $z=9.5$. Because our model follows a distinctly inside-out scenario, the filaments connecting the neutral regions are ionized first, producing isolated neutral islands. This contrasts with scenarios that involve a large degree of outside-in reionization (e.g. \citealt{finlator09}, see also \citealt{watkinson14,hutter19,pagano20}). Once neutral regions split off from the percolating cluster, they disappear quickly, i.e. they have low persistence. Particularly around $z=7$, many short-lived regions split off from the percolating cluster. In physical terms, once a neutral region is surrounded by the ionizing front on all sides, it is quickly ionized from the outside in. Unlike some other percolation problems, reionization is therefore asymmetrical. The first percolation, from the birth of the earliest bubbles to the formation of the percolating ionized cluster, takes about \SI{300}{\Myr} with many bubbles surviving for \SI{100}{\Myr} or more. The second percolation, from the breaking apart of the neutral cluster to the end of reionization, takes only \SI{60}{\Myr}.

Despite our finding that the number of enclosed neutral regions is at all times small, it would be misleading to say that there are no neutral patches. At redshift $z=6.5$, many neutral patches are considerable in size, occasionally \SI{5}{\Mpc} or more in radius. The topology is therefore one of rare but large neutral patches connected only by tenuous neutral tunnels, similar to the topology seen at the end of reionization in the model of \citet{kulkarni19}. This is also reflected visually in the slices of Fig.~\ref{fig:cosmic_web1}, discussed below.

\subsection{Percolation and filamentarity}\label{sec:perc_fil}

The sudden topological changes that occur during reionization can be understood using percolation theory \citep{furlanetto16a,bag18,pathak22}. A percolation transition occurs when an infinite cluster suddenly appears or disappears. In the case of reionization, two such transitions can be identified. The first occurs when enough ionized regions merge to form one connected structure from one side of the simulation box to the other. This happens at $z=9.4$ ($\xII=0.13$) in our fiducial `Faint Galaxies' model. The second percolation transition occurs when the neutral cluster breaks apart into smaller clusters, which occurs at $z=6.3$ ($\xII=0.92$). These are shown as vertical dotted lines in Fig.~\ref{fig:betti_tracks}, flanking the peak of the number of tunnels ($\beta_1$).

\begin{figure*}
	\centering
	\subfloat{
        \includegraphics{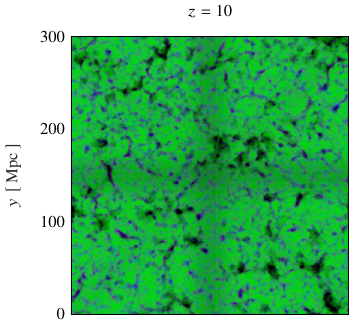}
	}
	\subfloat{
        \includegraphics{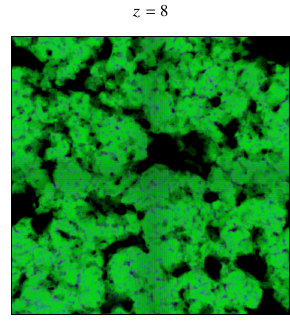}
	}
	\subfloat{
        \includegraphics{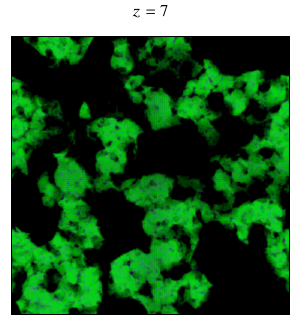}
	}\\\vspace{-1.5em}
	\subfloat{
        \includegraphics{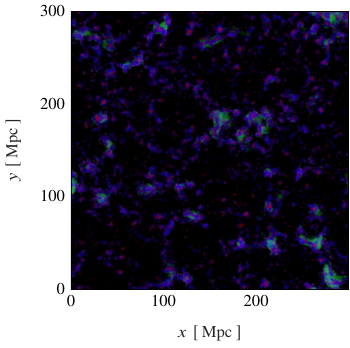}
	}
	\subfloat{
        \includegraphics{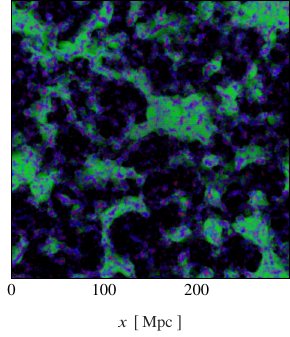}
	}
	\subfloat{
        \includegraphics{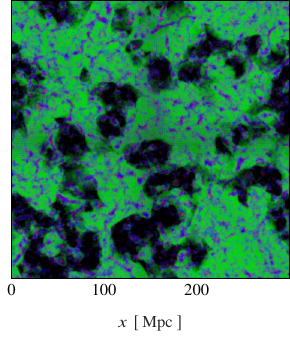}
	}
	\caption{Thin slices of the ionization topology for the `Faint Galaxies' model at $z=10$ (left), $z=8$ (middle), and $z=7$ (right). The neutral regions are shown at the top, coloured according to cosmic web signature: green = void, purple = wall, blue = filament, red = cluster. On the bottom, we show the ionized regions with the same colour scheme. Images created with the \texttt{Splotch} code \citep{dolag08}. The cube has side lengths of $\SI{300}{\Mpc}$.}
	\label{fig:cosmic_web1}
\end{figure*}

Topology and percolation theory are closely connected. In particular, the Euler characteristic $\chi$ can be used to detect percolation transitions. \citet{okun90} studied the Euler characteristic of uniformly distributed expanding balls, which is equivalent to the Poisson model considered in Paper I. He showed that
\begin{align}
\chi = \big(1-3x+x^2\big)e^{-x}, \label{eq:chi_curve}
\end{align}

\noindent
where $x(r)\sim r^3$ is the mean number of points in a ball of radius $r$. This functional form matches our results very well and shows an initial peak, followed by a valley, and a smaller second peak. These elements correspond to the bubble, tunnel, and patch stages that occur in sequence. Percolation transitions occur at the two zeros of $\chi$. In a wider range of models, a percolation transition occurs when $\chi\approx0$ \citep{neher08}, which happens when one topological feature starts dominating over another \citep{bobrowski20}. This also agrees with the broader class of models considered in Paper I, where we used the first percolation transition to define the beginning of the tunnel stage. Similar behaviour was reported by \citet{giri20} and broadly the same behaviour is seen again here, though with an important difference. For the `Faint Galaxies' model, the zero of $\chi$ occurs some time after the first percolation transition and for the `Bright Galaxies' model, $\chi$ never becomes negative at all. Rather than identifying the percolation transitions with the exact zeros of $\chi$, which does not hold in general, it seems more appropriate to associate percolation transitions with a rapid change in the number of tunnels. This connection can be understood by considering the structure of the percolating clusters.

Between $6.3<z<9.4$, there are two intertwined percolating clusters: one neutral and one ionized. In Fig.~\ref{fig:betti_tracks}, we see that this era corresponds to the period where tunnels are dominant. Furthermore, this is also the period during which the ionization fraction rises most rapidly, which we identify with the EoR proper. The dominance of the tunnels is not coincidental: $\beta_1$ represents the number of holes and therefore measures the degree to which these two clusters are entangled. This can be seen very clearly in Fig.~\ref{fig:3d_image} and the slices displayed in Fig.~\ref{fig:cosmic_web1}. The rapid change in filamentarity of the largest cluster is a hallmark of percolation \citep{bag18}. From the point of view of homology, this happens because higher-dimensional features are born when lower-dimensional features link together. The number of tunnels represents an important physical observable, related to the shape of the largest cluster \citep{pathak22}. For example, in the `Bright Galaxies' model, where the ionized regions surrounding the sources are larger, the ionized cluster contains fewer holes through which the neutral cluster could connect to itself. We thus see through the lens of percolation theory how the physics of reionization affects its topology.

\begin{figure*}
	\vspace{-1em}
	\subfloat{
	\includegraphics{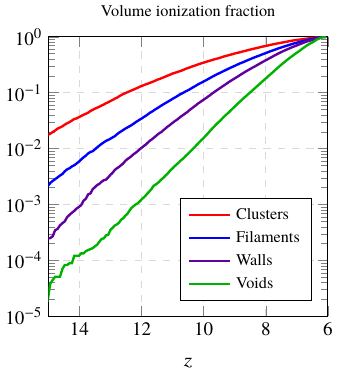}
	}
	\subfloat{
	\includegraphics{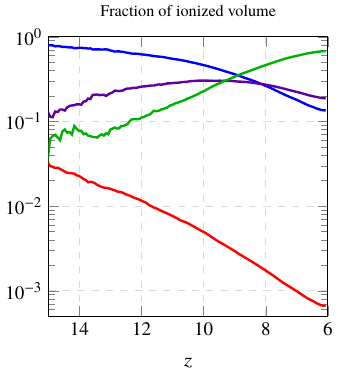}
	}
	\subfloat{
    \includegraphics{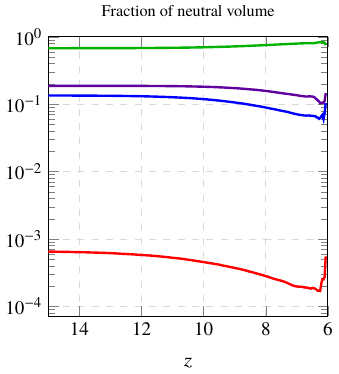}
	}

	\caption{(Left) Ionization fraction by cosmic web signature. (Middle \& Right) The fraction of ionized and neutral cells that is of a given signature. All three panels are for the `Faint Galaxies' model}
	\label{fig:web_signature}
\end{figure*}

\section{Cosmic Web Correspondence}\label{sec:cosmic_web}

The evolution of the ionization topology is intimately connected with the formation of the earliest structures that emerged during the Cosmic Dawn. These structures themselves have a rich topology, which reflects the hierarchical build-up of the cosmic web \citep{pranav16,wilding21}. In this section, we clarify the connection between topological features in the ionization bubble network on the one hand and the cosmic density field on the other. Both fields have topological features in dimensions $k=0,1,2$: the connected components, tunnels, and cavities of the respective fields. For the ionization bubble network, these correspond to ionized regions, ionized and neutral tunnels, and enclosed neutral patches. For the cosmic density field, they correspond to the well-known components of the cosmic web \citep{bond96b,weygaert08}: clusters, filaments, and cosmic voids. For inside-out models of reionization, such as the models studied here, there is a clear association between the $k$-dimensional components in the two fields. Ionizing sources are most likely to be found in the densest regions, such that most ionized regions initially expand outward from clusters. This establishes an initial correspondence between the zero-dimensional features in both fields. Similar reasoning applies to the filaments, which contain around $50\%$ of dark matter halos, this ratio varying only slightly over time \citep{cautun14}, and a similar fraction of galaxies \citep{ganeshaiah19}. Hence, filaments are the next most likely location for sources and once ionized connect the bubbles surrounding clusters. This provides a correspondence between the one-dimensional features: filaments and tunnels. Finally, due to a lack of sources, cosmic voids tend to be ionized much later, providing a connection with the two-dimensional neutral patches. While this basic correspondence appears to hold true, the details are considerably more complex as the following discussion reveals.

We describe the ionization of the various components of the cosmic web for our `Faint Galaxies' model. To identify the structural components of the cosmic web, we use the \nexus{}  algorithm \citep{cautun13} to calculate a `cosmic web signature' at each point in space. This signature
corresponds to one of the principal morphological elements that constitute the cosmic web: void, filament, wall or cluster. The \nexus{} formalism is the most commonly used version of the Multiscale Morphology Filter (MMF) and NEXUS families of cosmic web classification techniques \citep{aragon07,aragon10,cautun13,cautun14,aragon14}. A detailed description of this formalism is provided in Appendix \ref{sec:nexus}. Instrumental for these algorithms is that they simultaneously pay heed to two principal characteristics of the cosmic web. The first aspect concerns the mostly anisotropic components of the cosmic web, for the specification of which the algorithms invoke the eigenvalues of the Hessian of the density field, velocity field or tidal field.  Equally important is the \emph{Scale-Space} analysis used to probe the multiscale character of the cosmic mass distribution: the product of the hierarchical evolution and build-up of structure in the Universe. The outcome of the MMF/\Nexus{} identification procedure is a set of diverse and complex cosmic web components, from the prominent filamentary arteries to underdense cosmic voids. Amongst the various versions of \Nexus{} algorithms, we here use \nexus{}. This version uses a Log-Gaussian filtering of the cosmic density field as input. It is the version that is most used, due to its optimal dynamic range, resolving structural features of the cosmic web ranging from small tenuous features up to the large dominant arteries and voids of the Megaparsec Universe.

After identifying the components of the cosmic web by their signature, we analyse their corresponding ionization histories. In Fig.~\ref{fig:web_signature}, we show the volume ionization fraction $\xII(z)$ by cosmic web signature. In line with expectation, all four components follow a similar trajectory, but with filaments, walls, and voids (in that order) delayed behind the clusters. This ordering, in which the densest environments are ionized first, is the basic prediction for inside-out scenarios \citep{hutter17}. At high redshifts, filaments constitute most of the ionized volume. Even though clusters are ionized first, they make up a negligible fraction of the total volume. Meanwhile, the ionization front has not yet reached the walls and voids. This leaves the filaments as the dominant cosmic web environment for ionized material during the early stages of reionization. The bottom left panel of Fig.~\ref{fig:cosmic_web1} shows a rendering of the ionized regions at $z = 10$. The picture is clearly dominated by the blue filaments, which make up 50\% of the ionized volume at this time. At $z = 9$, filaments, walls, and voids each make up about a third of the ionized volume. During the EoR, the voids are also ionized and they take over as the dominant component due to their larger total volume.

Let us consider next the identity of the neutral patches. For inside-out scenarios, the neutral patches are expected to coincide with the deepest voids in the cosmic web. This expectation is confirmed visually by the top row of Fig.~\ref{fig:cosmic_web1}, showing renderings of the neutral regions at $z\in\{10,8,7\}$. The final snapshot shows the large remaining neutral patches entirely colour-coded as void regions. In the third panel of Fig.~\ref{fig:web_signature}, we see that voids constitute most of the neutral volume at all times, with the ratio increasing from $z=10$ onward as the last neutral walls and filaments are ionized. One interesting implication is that counting the tunnels in the neutral regions gives a lower bound on the filamentarity of the cosmic web.

As we saw in Section \ref{sec:evolution}, the tunnels are the most interesting and prominent feature of the ionization bubble network during reionization. The question arises whether they are related to the filaments of the cosmic web. At this point, we should be more explicit about what we mean by tunnels. Because the neutral and ionized regions are exactly complementary, the number of one-dimensional holes in the neutral regions (or `ionized tunnels') is equal to the number of one-dimensional holes in the ionized regions (or `neutral tunnels'). Of course, this number is just what $\beta_1$ measures. This result applies only to the one-dimensional holes and is a consequence of Alexander duality (see \citealt{hatcher01}). Conveniently, this means that we can use both the neutral and the ionized tunnels to constrain $\beta_1$. In the `Faint Galaxies' simulations, the filaments are ionized early on. Hence, the tunnels that connect the ionized regions most likely coincide with these filaments. On the other hand, the neutral tunnels can be found even in the deepest void regions.

\begin{figure*}
    \subfloat{
		\includegraphics{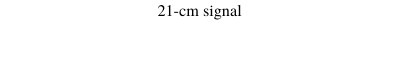}
	}
    \subfloat{
	    \includegraphics{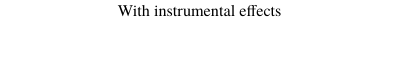}
	}~\\\vspace{-3em}
	\subfloat{
		\includegraphics{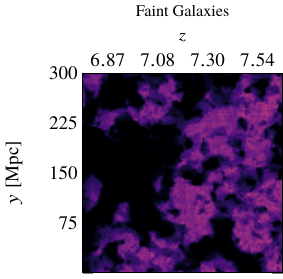}
	}
	\subfloat{
		\includegraphics{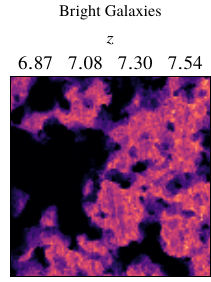}
	}
	\subfloat{
		\includegraphics{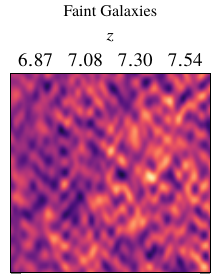}
	}
	\subfloat{
		\includegraphics{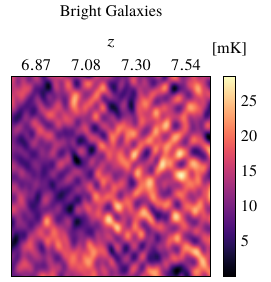}
	}~\\\vspace{-1em}
	\hspace{0.5em}\subfloat{
	    \includegraphics{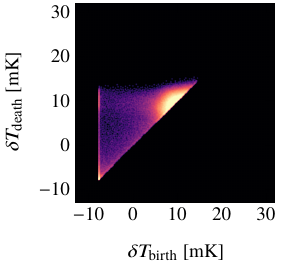}
	}\hspace{-0.5em}
	\subfloat{
		\includegraphics{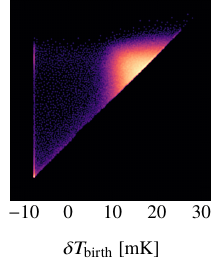}
	}\hspace{-0.5em}
	\subfloat{
		\includegraphics{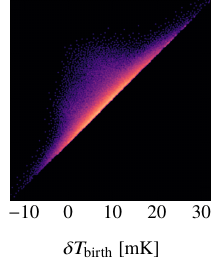}
	}\hspace{-0.5em}
	\subfloat{
		\includegraphics{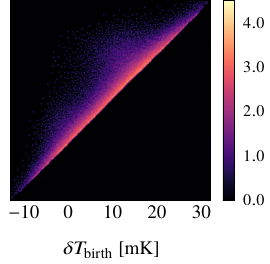}
	}
	\caption{(Top) Slices through the theoretical lightcones for the `Faint Galaxies' and `Bright Galaxies' models before and after applying telescope effects, assuming the optimistic foreground scenario with $10^4$ hours of observing. (Bottom) One-dimensional persistence fields averaged over 16 realizations.}
	\label{fig:thermal_filtration}
\end{figure*}

\section{Classification}\label{sec:thermal}

The predicted differential brightness temperature $\delta T_b$ is given by Eqn. \eqref{eq:delta_T}. In principle, we can extract the ionized and neutral regions from the observed signal using a variety of techniques such as granulometry \citep{kakiichi17} or the friends-of-friends algorithm \citep{friedrich11}. This paves the way for an analysis of the type described above. The approach in this case is very similar to the ones needed to extract other statistics such as the size distribution of the ionized or neutral regions. However, another interesting possibility is to apply our formalism directly to the extracted temperature field itself by means of a thermal filtration.

\begin{figure*}
	\normalsize
	\subfloat{
		\includegraphics{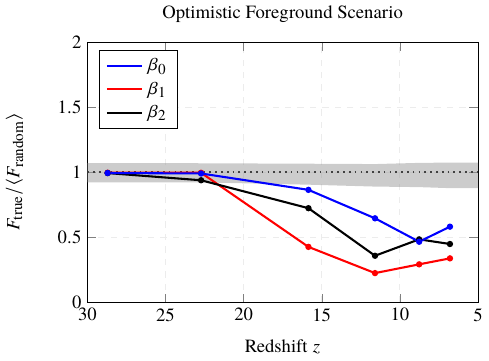}
	}
	\subfloat{
	   \includegraphics{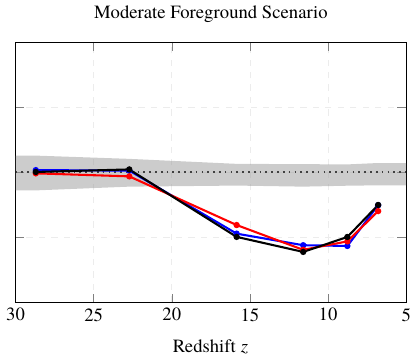}
	}
    \caption{Ratio of the test statistic $F$ evaluated on true classifications of mock observations compared to the average value for random permutations of the labels. The shaded area is the 95\% CI for random permutations. The fact that $F_\text{true}$ lies far outside this region, indicates that observations can be differentiated on the basis of Wasserstein distances between persistence diagrams. The ratio is shown as a function of redshift, for all dimensions and both foreground scenarios, assuming the fiducial number of $10^3$ hours of observation. }
	\label{fig:distance_betti}
\end{figure*}

\begin{figure*}
	\normalsize
	\subfloat{
		\includegraphics{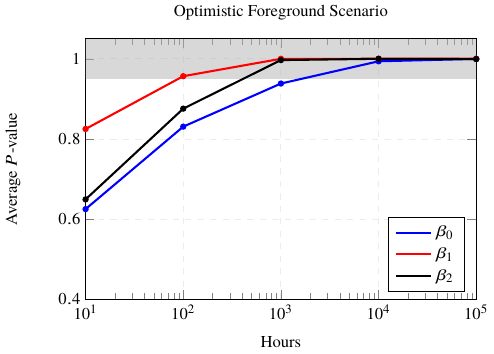}
	}
	\subfloat{
	   \includegraphics{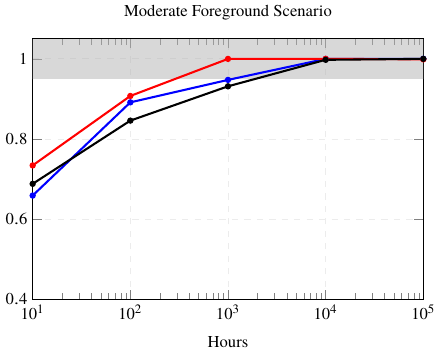}
	}
    \caption{Average $P$-value assuming that observations from the `Faint Galaxies' and `Bright Galaxies' models are topologically indistinguishable on the basis of $\beta_d$-persistence diagrams ($d=0,1,2$), shown as a function of hours of observation and for different foreground scenarios. The results are averaged over the tomographic slices with $z\leq17$. The shaded area represents $P>0.95$. The constraining power of the tunnels ($\beta_1$) is greatest under the optimistic scenario and comparable to the other features under the moderate scenario.}
	\label{fig:accuracy_dimensions}
\end{figure*}

\subsection{Thermal Filtrations}

One of the main advantages of the persistent homology framework is its ability to distinguish real topological features from noise \citep{edelsbrunner00}. Small perturbations in the signal result in persistence diagrams that are close to the unperturbed diagram. This follows from the stability of the Wasserstein metric \eqref{eq:wasserstein}. Specifically, this is the case because because small perturbations either change the persistence of real features by a small amount or create new noisy features with small persistence. In addition, the preceding analysis has shown that persistent homology is sensitive to tunnels and that these are a key tracer of the percolation processes associated with reionization. Such tunnels should also be present in the temperature field itself. This reasoning provides strong motivation for applying the formalism directly to the temperature field of the observed 21-cm signal. This can be done with the Field Filtration method outlined in Section \ref{sec:methods}. The output is a thermal filtration of the signal, revealing which topological features emerge or disappear as the temperature is increased.

We demonstrate this in Fig.~\ref{fig:thermal_filtration} for a tomographic slice of the $\delta T_b$ field at $z = 7$ from a lightcone created with \TOcmFast. For illustrative purposes, we assume the optimistic foreground scenario and decrease the fiducial noise level by a factor $\sqrt{10}$. The top row in Fig.~\ref{fig:thermal_filtration} shows images of the signal, before and after applying telescope effects, for the two astrophysical source models: `Faint Galaxies' and `Bright Galaxies'. The noiseless images reveal cold spots within hot regions, the most prominent of which survive in the noisy images, and which correspond to ionized cavities or tunnels that protrude through neutral regions. We also see temperature fluctuations within hot regions that arise as a result of variations in the baryonic overdensity, spin temperature, or ionization fraction, according to Eqn. \eqref{eq:delta_T}. The bottom row shows the associated $\beta_1$-persistence diagrams for tunnels. There are two notable classes of highly persistent features in the noiseless diagrams, one narrow strip at $\delta T_b=\SI{-10}{\milli\K}$ (keeping in mind that the absolute temperature is arbitrary) and one broad grouping at a higher temperature that depends on the model. The first class corresponds to tunnels in the ionized network, which can be identified with the tunnels studied in the previous sections. The second class corresponds to tunnels in the signal that emerge as a result of temperature fluctuations within neutral regions. A third intermediate class of highly persistent features born at temperatures $-10 < \delta T_b < \SI{10}{\milli\K}$ may be identified with partially ionized tunnels. After applying the telescope effects, the two main classes can still be differentiated by eye, especially for the `Bright Galaxies' model. Under more pessimistic observational circumstances, when differences are harder to see by eye, a statistical approach based on distances between persistence diagrams may still allow an interpretable topological analysis. This will be the topic of the remainder of this section.

\subsection{Model selection}

We have seen that the `Faint Galaxies' and `Bright Galaxies' models produce distinct ionization topologies. Let us consider whether these differences are statistically significant for noisy observations under different observational circumstances. Using the pipeline discussed in Section \ref{sec:sims}, we generate a catalogue of mock lightcones for both models. Each lightcone consists of a set of three-dimensional tomographic `slices' and for each slice, we construct a thermal filtration and compute the associated persistence diagrams for $\beta_0,\beta_1,\beta_2$. The corresponding persistence diagrams from different lightcones generated with the same model are statistically independent and identically distributed. We imagine that these might correspond to non-overlapping fields observed during each night.

To determine whether the differences are statistically significant, we will carry out a randomization test. Given a set of $N$ persistence diagrams, some of which were generated with one model and some with the other, our task is to label the diagrams according to the underlying model, dividing them into two disjoint sets of $n_1$ and $n_2=N-n_1$ elements. Following \cite{robinson17}, we use the test statistic
\begin{align}
F = \sum_{m=1}^2 \frac{1}{2n_m(n_m-1)}\sum_{j=1}^{n_m}\sum_{k=1}^{n_m}d(X_{m,j},X_{m,k})^2, \label{eq:test_F}
\end{align}

\noindent
where $d(X,Y)$ is the Wasserstein metric defined in Eqn.~\eqref{eq:wasserstein} and $X_{m,j}$ is the $j$th persistence diagram labelled with $m$ and $m\in\{1,2\}$ is arbitrary. Note that there are multiple diagrams per lightcone and we only compute distances between the corresponding diagrams from different lightcones. Our results are based on $N=16$ lightcones, half from each model. $F$ is the mean squared distance between diagrams \emph{with the same label}, based on the observation that Wasserstein distances are minimized for pairs of diagrams from the same model (see Fig.~\ref{fig:distance_example}). This statistic is relatively cheap to compute compared to statistics that involve the Fr\'echet average or cross distances. Let $F_\text{true}$ be the value of the test statistic evaluated in the case where each observation is labelled correctly (all Faint observations are in one group and all Bright observations in the other). We may reject the hypothesis that the two models are topologically indistinguishable if $F_\text{true}$ is extreme compared to the value of $F$ for random permutations of the labels.

As an illustration, Fig.~\ref{fig:distance_betti} shows the ratio $F_\text{true}/\langle F_\text{random}\rangle$ as a function of redshift, assuming the fiducial number of $10^3$ hours of observation. The ratio is most extreme for the optimistic foreground scenario (left panel) and lies far outside the shaded region indicating the 95\% range for random permutations. This shows that topological differences are significant even for noisy observations. As anticipated, the tunnels ($\beta_1$) are the strongest differentiator throughout. We also recognize that the cavities $(\beta_2)$ are a stronger indicator at early times, while the components $(\beta_0)$ are relatively more discriminating at late times. Although these results are harder to interpret than the noiseless ionization fields studied in the previous sections, this may be understood by noting that a class of cavities in the temperature field correspond to ionized regions. These differ more strongly between the two models at early times, because both models are tuned to reproduce the same global reionization history at late times (see Fig.~\ref{fig:sources}). By contrast, a class of components $(\beta_0)$ of the temperature field correspond to neutral regions, which are more sensitive to the model at late times. Under the moderate foreground scenario (right panel), all features exhibit the same redshift dependence, but we see hints of the same pattern.

Using a Monte Carlo approach, we estimate the probability $P(F_\text{true}\leq F_\text{random})$ to determine whether $F_\text{true}$ is extreme and hence whether the topological differences are significant under different observational circumstances. Fig.~\ref{fig:accuracy_dimensions} shows the resulting $P$-values as a function of the number of hours of observation, averaged over the tomographic slices with $z\leq 17$. We see that the distinguishing power of the tunnels $(\beta_1)$ is greatest in the case of the optimistic foreground scenario, and similar to the other features in the moderate scenario. Focussing on the tunnels, the differences are extreme for the fiducial number of hours ($10^3$) and greater in both foreground scenarios. For $\SI{e2}{\hrs}$, we still obtain $P>0.95$ for the optimistic scenario and $P>0.90$ for the moderate scenario. These results suggest that persistence diagrams from thermal filtrations can be used to extract astrophysical information from noisy observations.

\begin{figure}
	\normalsize
	\subfloat{
		\includegraphics{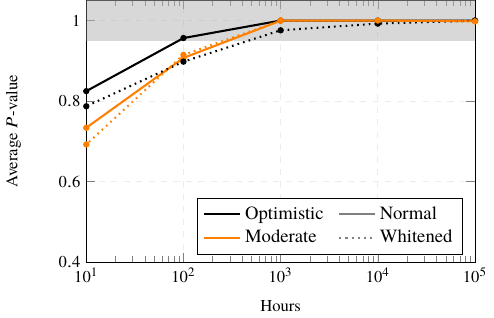}
	}
    \caption{Average $P$-value assuming that observations from the `Faint Galaxies' and `Bright Galaxies' models are topologically indistinguishable on the basis of $\beta_1$-persistence diagrams (tunnels), as a function of hours of observation for different foreground scenarios. The results are averaged over the tomographic slices with $z\leq17$. Also shown as dotted lines are the results obtained from whitened temperature maps, which by construction have uninformative power spectra. The shaded area represents $P>0.95$.}
	\label{fig:accuracy}
\end{figure}

Topological approaches rely on both amplitude and phase information and can therefore offer discriminatory power beyond what is possible with two-point statistics alone. To demonstrate this explicitly, we apply our pipeline to whitened temperature fields, obtained by dividing out all the information contained in the three-dimensional power spectrum. We define the whitened temperature map in Fourier space, $\epsilon(\mathbf{k})$, by
\begin{align}
	\epsilon(\mathbf{k}) = \frac{T(\mathbf{k})}{\sqrt{\langle\rvert T(k)\rvert^2\rangle}},
\end{align}

\noindent
where we take into account the Fourier space masking used to model the instrumental effects (Section \ref{sec:instrumentals}). By construction, the power spectrum of $\epsilon(\mathbf{k})$ is completely uninformative: $P(k)=\langle\rvert\epsilon(k)\rvert^2\rangle=1$. However, the fields still contain amplitude information, as we do not enforce $\rvert\epsilon(\mathbf{k})\rvert=1$. The dotted lines in Fig.~\ref{fig:accuracy} show the $P$-values obtained for these whitened fields, using the $\beta_1$-persistence diagrams. The results are slightly degraded compared to the normal temperature maps, but topological classification still appears to be possible for the fiducial number of hours ($10^3$) and greater. This explicitly confirms that persistent homology is complementary to the 21-cm power spectrum.

\section{Discussion}\label{sec:discussion}

Among the physical processes studied in cosmology, reionization is particularly well-suited for a topological description. As the last major phase transition of gas in the Universe, reionization describes the process by which the neutral hydrogen of the Dark Ages was transformed into the ionized intergalactic medium (IGM) seen today. From the point of view of topology, the evolution of the IGM during the Epoch of Reionization is characterized by the spatial connectivity of the ionized and neutral regions. In this work, we have analysed this connectivity using the language of persistent Betti numbers. Borrowed from algebraic topology, the $k$th Betti number $\beta_k$ represents the number of $k$-dimensional holes in an object, or formally the rank of the $k$th homology group \citep{hatcher01,edelsbrunner10b,carlsson21}. For three-dimensional structures like the ionization bubble network, there are three non-trivial Betti numbers: $\beta_0$ gives the number of connected components, $\beta_1$ the number of one-dimensional openings or tunnels, and $\beta_2$ the number of cavities or shells. In the context of reionization, these are interpreted as ionized bubbles $(\beta_0)$, ionized or neutral tunnels $(\beta_1)$, and enclosed neutral islands $(\beta_2)$. We collectively refer to these $k$-dimensional holes as topological features. By following the births and deaths of features as a function of a filtration parameter $\alpha$, we construct persistence intervals $(\alpha_\text{birth},\alpha_\text{death})$, describing the range of values for which the features exist \citep{edelsbrunner00,zomorodian05}. Combining the persistence intervals of all features yields a persistence diagram, capturing at once the topological evolution as a function of time, scale or temperature. One of the key advantages of the framework is that the measured quantities can be linked directly to morphological features, including the prominent tunnels of the bubble network. The framework also allows the identification of the most significant features present in noisy observations.

Using the language of persistent homology, we studied the evolution of the ionization topology for semi-numerical models generated with the \TOcmFast{} code \citep{murray20}. Different stages of reionization can be identified by the types of features that dominate. When the first stars ionize the hydrogen around them, the ionized bubbles initially trace the large-scale topology of the cosmic web. Between $5.5\lesssim z\lesssim 10$, an intricate network of ionized and neutral regions emerges, whose topology depends both on the location and properties of the sources and sinks of ionizing radiation. The tunnels that connect the ionized patches are the dominant component during this period and depend most sensitively on the physics of reionization (see Fig.~\ref{fig:sources}). We find that this tunnel stage coincides with the period during which the ionization fraction rises most rapidly, which can be associated with the Epoch of Reionization proper, and which begins and ends with a percolation transition \citep{furlanetto16a,bag18}. In the first percolation transition, the ionized bubbles link up to form an infinite percolating cluster of ionized material, perforated by a multitude of neutral tunnels. In the second (de-)percolation transition, the infinite cluster of neutral material splits apart into many disconnected patches, while the neutral tunnels are ionized and disappear. Unlike some idealized percolation models, we find an asymmetry between the two transitions, with the first stage of reionization lasting much longer than the final stage and with ionized bubbles being much more persistent (long-lived) and numerous compared to the neutral islands at their respective peaks.

We also explored the relation between the morphological components of the ionization bubble network (bubbles, tunnels, and islands) and those of the cosmic web (clusters, filaments, and voids), finding a close association between the $k$-dimensional features of both, particularly between the filaments of the cosmic web and the tunnels of ionized material during the early stages of reionization. Finally, we applied our formalism directly to mock observations of the 21-cm signal, assuming \SI{1000}{\hrs} of observation with Phase II of HERA \citep{deboer17}, for different foreground scenarios \citep{pober14}. By casting the signal in terms of persistence diagrams using the temperature itself as filtration parameter, significant topological features can be differentiated from thermal fluctuations. We used the $L^2$-Wasserstein metric as a topological distance measure between persistence diagrams \citep{turner14,boisonnat18}. Applying a randomization test to these Wasserstein distances \citep{robinson17}, we showed that observations from different astrophysical models are distinguishable, even for whitened temperature fields which have the information content of the power spectrum divided out. To use persistent homology for astrophysical parameter inference, observations will have to be compared with a large number of theoretical models. Although this is possible with Eqn.~\eqref{eq:test_F}, an alternative would be to use a kernel density-based likelihood function \citep{mileyko11} or to use vector representations of persistence diagrams like persistence fields (Paper I), persistence images \citep{adams17,cole20} or persistent Betti functions \citep{heydenreich20}. Markov Chain Monte Carlo methods could then be used to extract parameter constraints, analogous to approaches for the 21-cm power spectrum \citep{greig15} and bispectrum \citep{watkinson22}.

In an effort to find complementary observables to the power spectrum, we have identified persistence diagrams as a sensitive probe of the ionization topology. They can be related to several other quantities that have been used to study reionization. First of all, the Betti numbers $\beta_i(\alpha)$ for $i\in\{0,1,2\}$ count the number of features alive as a function of $\alpha$ (\citealt{elbers17,whe_2018a,kapahtia18,kapahtia19,kapahtia21,giri20,bianco21}) and are therefore an `integral' of the persistence diagrams. The commonly used Euler characteristic \citep{lee08,friedrich11,hong14} is an alternating sum of Betti numbers, $\chi=\beta_0-\beta_1+\beta_2$, and one of the Minkowski functionals \citep{gleser06,yoshiura16,kapahtia18,bag18,chen19}. These quantities are frequently shown as a function of a course-graining scale $\alpha$. For Gaussian random fields, the shape of the resulting Betti curves is sensitive to the power spectrum. This is unlike the equivalent $\chi(\alpha)$-curve, which only depends on the power spectrum in its overall amplitude \citep{pranav19b}. Interestingly for reionization, these topological quantities are also sensitive to percolation transitions \citep{furlanetto16a,bag18}. Indeed in an idealized setting, $\chi$ passes through zero at a percolation transition, indicating that one component starts to dominate over another \citep{neher08}. The twin percolation transitions of reionization are shown even more clearly in the $\beta_1$-persistence diagram for tunnels (Fig.~\ref{fig:iof_results}, middle panel), where the edges of the triangle at $x_{\text{II},\text{birth}}\sim0.1$ and $x_{\text{II},\text{death}}\sim0.95$ reflect the onset of the tunnel stage at the first percolation transition and the disappearance of tunnels at the second. Persistence diagrams can also be represented as merger trees \citep{chardin12}, showing the lifetimes of features (Fig.~\ref{fig:merger_tree}). Using a spatial filtration, persistent homology allows a multiscale study of topological features (Paper I), relating persistence diagrams to bubble and island size distributions \citep{lin16,giri17,kakiichi17,bag18}.

In Paper I, our analysis was restricted to phenomenological models of reionization. Although the resulting idealized bubble networks are topologically isomorphic to networks with more complex morphologies, thereby providing a connection with theoretical results such as Eqn.~\eqref{eq:chi_curve}, this nevertheless limited the applicability of our results. In this paper, we have expanded our analysis to semi-numerical models generated with \TOcmFast{}. Compared to Paper I, we find broadly similar results in terms of the topological stages of reionization, the role of the percolation transitions, the rapid rise of the ionization fraction during the tunnel stage, and the asymmetry between the ionized regions and neutral patches at their respective peaks. These appear to be generic features of bubble reionization scenarios, although in detail the topology retains a strong dependence on the underlying physics. By varying the parameters of the model, we explored some of this dependence. Of course, the fidelity of the simulations could be further improved. At higher resolutions, smaller topological features could be identified, which may be important for the rare neutral islands. The semi-numerical models could be extended to include other relevant processes, such as redshift space distortions \citep{bharadwaj04,barkana05,mao12}, relative baryon-dark matter velocities \citep{tseliakhovich10,dalal10}, and molecular-cooling galaxies in minihalos \citep{qin20,munoz22}. Finally, we expect that the application of persistent homology to self-consistent radiation hydrodynamics simulations \citep{gnedin14b,rosdahl18,ocvirk20,chan21,kannan22} will offer further insights.

\section*{Acknowledgements}

We thank the anonymous referee for useful comments that helped improve the manuscript. RvdW is grateful for numerous useful, instructive, and insightful discussions with Gert Vegter, Bernard Jones, Job Feldbrugge, Garrelt Mellema, Keimpe Nevenzeel, Marius Cautun, Pratyush Pranav, and Georg Wilding. WE thanks Carlos Frenk, Adrian Jenkins, Baojiu Li, and Silvia Pascoli for valuable discussions and encouragement. WE is supported by the Durham Prize Scholarship in Astroparticle Physics and the UK Science and Technology Facilities Council Consolidated Grants No. ST/P000541/1 and ST/T000244/1. This work used the DiRAC@Durham facility managed by the Institute for Computational Cosmology on behalf of the STFC DiRAC HPC Facility (www.dirac.ac.uk). The equipment was funded by BEIS capital funding via STFC capital grants ST/K00042X/1, ST/P002293/1, ST/R002371/1 and ST/S002502/1, Durham University and STFC operations grant ST/R000832/1. DiRAC is part of the National e-Infrastructure.

%%%%%%%%%%%%%%%%%%%%%%%%%%%%%%%%%%%%%%%%%%%%%%%%%%
\section*{Data availability}

The software packages \TOcmFast{} and \TOcmSense{} are publicly available. The code used for topological calculations is available at \url{http://willemelbers.com/persistent-homology/}. Our pipeline for including telescope effects and calculating topological distances is available at \url{https://github.com/wullm/21runs}. The generated mock observations will be made available upon request to the corresponding author.

%%%%%%%%%%%%%%%%%%%% REFERENCES %%%%%%%%%%%%%%%%%%

% The best way to enter references is to use BibTeX:

\bibliographystyle{mnras}
\bibliography{main} % if your bibtex file is called main.bib

%%%%%%%%%%%%%%%%%%%%%%%%%%%%%%%%%%%%%%%%%%%%%%%%%%

%%%%%%%%%%%%%%%%% APPENDICES %%%%%%%%%%%%%%%%%%%%%

\appendix

\section{The MMF/Nexus Formalism}\label{sec:nexus}

The family of Multiscale Morphology Filter (MMF) and NEXUS techniques \citep{aragon07,aragon10,cautun13,cautun14,aragon14} are used for the morphological identification of cosmic web structures. A key aspect of these approaches is the use of a \emph{Scale-Space} representation that ensures the detection of structures present at all scales. The formalism entails a fully adaptive framework for classifying the matter distribution on the basis of local variations in the density field, velocity field or gravity field. At each scale, these are encoded in the Hessian matrix of the field in question. Subsequently, a set of morphological filters is used to classify the spatial matter distribution into three basic components: the clusters, filaments, and walls that constitute the cosmic web, with cosmic voids assuming the remaining space. Each volume element is assigned a single environmental characteristic by requiring that filament regions cannot be nodes and that wall regions can be neither nodes nor filaments. The outcome of the identification procedure is a volume-filling field that specifies at each point the local morphological signature: node, filament, wall or void.

\subsection{Formalism}
The first step of MMF/NEXUS is to define a four-dimensional scale-space representation of the input field $f(\mathbf{x})$. In nearly all implementations, this is achieved by means of a Gaussian filtering of $f(\mathbf{x})$ over a set of scales $[R_0, R_1,\dots,R_N]$, given by
\begin{align}
	f_{R_n}(\mathbf{x}) = \int\frac{\mathrm{d}^3k}{(2\pi)^3}e^{-k^2R_n^2/2}\hat{f}(\mathbf{k})e^{i\mathbf{k}\cdot \mathbf{x}},
\end{align}

\noindent
where $\hat{f}(\mathbf{k})$ is the Fourier transform of $f(\mathbf{x})$. Subsequently, the Hessian $H_{ij,R_n}(\mathbf{x})$ of the filtered field is computed as
\begin{align}
	H_{ij,R_n}(\mathbf{x}) = R_n^2\frac{\partial^2 f_{R_n}(\mathbf{x})}{\partial x_i\partial x_j}.
\end{align}

\noindent
In this equation, $R_n^2$ serves as a renormalization factor related to the multiscale nature of the algorithm. The morphological signature is contained in the local geometry as specified by the eigenvalues of the Hessian matrix, $h_1 \le h_2 \le h_3$. The eigenvalues are used to assign to every point, $\mathbf{x}$, a node, filament, and wall characteristic, which are determined by a set of morphology filter functions (see \citealt{aragon07,cautun13}). The morphology filter operation consists of assigning to each volume element and at each filter scale an environmental signature $\mathcal{S}_{R_n}(\mathbf{x})$. Subsequently, the environmental signatures calculated for each filter scale are combined to obtain a scale-independent signature, $\mathcal{S}(\mathbf{x})$, which is defined as the the maximum signature over all scales,
\begin{align}
    \mathcal{S}(\mathbf{x}) = \max_{\rmn{levels\;}n} \mathcal{S}_{R_n}(\mathbf{x})\,. \label{eq:total_response}
\end{align}

\noindent
The final step in the MMF/\Nexus{} procedure involves the use of criteria to find the threshold signature that identifies valid structures. Signature values larger than the threshold correspond to real structures, while the remainder are regarded as spurious detections. The various implementations of MMF/NEXUS can differ in the definition of the detection thresholds.

\subsection{Application and Developments}

Following the introduction of the basic version of MMF by \citet{aragon07}, the technique was applied to the analysis of the cosmic web in simulations of cosmic structure formation \citep{aragon10} and to the identification of filaments and galaxy-filament alignments in the SDSS galaxy distribution \citep{jones10}. The principal technique, and corresponding philosophy, has subsequently been branched in several further elaborations and developments. The two principal developments are the NEXUS formalism  developed by \citet{cautun13} and the MMF-2 method developed by \citet{aragon14}. NEXUS has extended the MMF formalism to a substantially wider range of physical agents involved in the formation of the cosmic web, along with substantially firmer foundations for the criteria used to identify the various weblike structures. MMF-2 instead focusses on the hierarchical nature of the cosmic web, by introducing and exploiting the concept of \emph{hierachical space}.

\subsubsection{\Nexus{} and \nexus{}}
The \Nexus{} version of the formalism \citep{cautun13,cautun14} builds upon the original MMF algorithm and was developed with the goal of obtaining a more robust and more physically motivated environmental classification method. The full \Nexus{} suite of cosmic web identifiers (see \citealt{cautun13}) includes options for a range of cosmic web tracers, such as the ordinary density, the logarithmic density, the velocity divergence, the velocity shear, and the tidal force fields. \Nexus{} has incorporated these
options in a versatile code for the analysis of cosmic web structure and dynamics, following the realization that they represent key physical aspects that shape the cosmic mass distribution into the complexity of the cosmic web.

Amongst the various versions of the \Nexus{} suite, we made use of \nexus{} in this paper. This is the version that is most used, due its considerable dynamic range. Other versions of \Nexus{}, particularly those looking at the anisotropy of the velocity field, tend to single out the dynamically dominant features (see e.g. \citealt{ganeshaiah18}). \nexus{} takes as input a regularly sampled density field, which is smoothed using a \logFilter{} filter. Like the basic version of the formalism, the filter is applied over a set of scales and for each scale, the eigenvalues of the Hessian matrix are computed. The eigenvalues subsequently define an environmental signature for each volume element that characterizes how close this region is to an ideal knot, filament, and wall. Then, the environmental signatures computed for each scale are combined into a single scale-independent signature. In the last step, physical criteria are used to determine a detection threshold. All points with signature values above the threshold are valid structures. For knots, the threshold is given by the requirement that most knot-regions should be virialized. For filaments and walls, the threshold is determined on the basis of the change in filament and wall mass as a function of signature. The peak of the mass variation with signature delineates the most prominent filamentary and wall features of the cosmic web.

\subsubsection{MMF-2: Multiscale Morphology Filter-2} \label{section:MMF}

An alternative development of the original Multiscale Morphology Filter method is MMF-2 \citep{aragon14}. In order to account for the hierarchical nature of the cosmic web, MMF-2 introduces the concept of \textit{hierarchical space}. This is in contrast to scale-space approaches (as in the original MMF), which emphasize the scale of the structures but are insensitive to their nesting relations. The first step is the creation of a hierarchical space \citep{aragon10,aragon14}. This is done by Gaussian-smoothing the initial conditions (instead of the final density field). This linear-regime smoothing is applied when the Fourier modes are independent, such that specific scales in the density field can be targetted before Fourier mode-mixing occurs. When evolved under gravity, the smoothed initial conditions produce all the anisotropic features of the cosmic web, but lack small-scale structures below the smoothing scale. This reduces the dynamic range in the density field and greatly limits the contamination produced by dense halos in the identification of filaments and walls. The hierarchical space is a continuum covering the full range of scales in the density field. For practical purposes however, only a small set of linear-regime smoothed initial condition are evolved to the present time.

For each realization in the hierarchical space, a number of morphology filters are applied, defined by ratios between the eigenvalues of the Hessian matrix. Similar to other versions of MMF/NEXUS, a threshold is applied to the response from each morphology filter to produce a set of binary masks sampled on a regular grid that indicates which voxels belong to a given morphology at a given hierarchical level.

%%%%%%%%%%%%%%%%%%%%%%%%%%%%%%%%%%%%%%%%%%%%%%%%%%

% Don't change these lines
\bsp	% typesetting comment
\label{lastpage}
\end{document}